\documentclass{article}
\usepackage{frascatiphys}
\usepackage{graphicx}
\usepackage{wrapfig}
\usepackage[
colorlinks=true
,urlcolor=blue
,anchorcolor=blue
,citecolor=blue
,filecolor=blue
,linkcolor=blue
,menucolor=blue
,linktocpage=true
,pdfproducer=medialab
,pdfa=true
]{hyperref}
\usepackage{cite}
\usepackage{amsmath,amssymb}
\begin{document}
\title{ 
Charged Lepton Flavor Violation: A Theory Overview
}
\author{
P. S. Bhupal Dev       \\
{\em Department of Physics and McDonnell Center for the Space Sciences,} \\
{\em Washington University, St.~Louis, MO 63130, USA} 
}
\maketitle
\baselineskip=11.6pt
\begin{abstract}
We present a theory-oriented overview of the significance of charged-lepton flavor violation -- one of the cleanest windows onto physics beyond the Standard Model. We also summarize the current experimental landscape and illustrate how different measurements can diagnose the underlying new physics. 
\end{abstract}
\baselineskip=14pt

\section{Introduction}
The Standard Model (SM) has an accidental flavor symmetry $U(1)_e\times U(1)_\mu\times U(1)_\tau$. This leads to the perturbative conservation of lepton flavor, as well as conservation of total lepton number. Therefore, any observation of charged lepton flavor violation (CLFV) would be an unambiguous signal of physics beyond the SM (BSM)~\cite{Bernstein:2013hba, Calibbi:2017uvl, Ardu:2022sbt, Davidson:2022jai}.

The historical origin of this viewpoint is closely connected with the observation of muon decay and the recognition that the muon behaves as a distinct lepton flavor. 
From a modern perspective, neutrino oscillations demonstrate that lepton flavor is not a fundamental exact symmetry. Yet, despite the established violation of flavor in the neutrino sector, no charged-lepton flavor violating transitions, such as $\mu\to e\gamma$, $\mu\to3e$,  have been observed.

This contrast is not an accident. In the minimal SM extension with light neutrino masses, CLFV is generated only through loops containing neutrinos. The corresponding amplitudes are suppressed by the GIM mechanism and by ratios of neutrino masses to the electroweak scale~\cite{Petcov:1976ff}. For $\mu\to e\gamma$, for instance, the resulting branching ratio (BR) is of order $10^{-54}$ for the observed neutrino mass scale, many orders of magnitude below any realistic experimental sensitivity. This is the key reason CLFV is often described as a ``smoking gun'' for new physics. A useful way of seeing the exceptional sensitivity is to compare the neutrino contribution with a generic heavy scale $\Lambda$. A new interaction can replace the tiny suppression $m_\nu^2/m_W^2$ by a factor schematically of order $m_F^2/\Lambda^2$, where $m_F$ is a mass associated with the new particles participating in the CLFV interaction and $\Lambda$ is the corresponding new physics scale. Consequently, CLFV rates in BSM scenarios can be enhanced by tens of orders of magnitude while remaining consistent with the smallness of neutrino masses. In fact, many BSM mechanisms proposed to address other problems, such as neutrino mass, baryogenesis, dark matter, the flavor hierarchy, or anomalous magnetic moments, also generate CLFV potentially observable in current/future experiments, thus making it a promising tool.

The absence of a sizable SM background is a major experimental advantage of CLFV searches. Quark flavor-changing processes, by contrast, are allowed in the SM via the CKM mixing and are routinely observed. For example, BR($b\to s\gamma$) is at the $10^{-4}$ level~\cite{Misiak:2015xwa} and is an irreducible background for many BSM searches. The corresponding charged-lepton channels have no analogous irreducible SM contribution of observable size. Thus a convincing observation of CLFV would be an unambiguous indication of BSM physics. 

The experimental and theoretical literature on CLFV is extensive; useful broad reviews include Refs.~\cite{Bernstein:2013hba, Calibbi:2017uvl, Ardu:2022sbt, Davidson:2022jai}. The purpose of this proceedings contribution, based on the theory overview talk given at CLFV 2026~\cite{clfv5}, is not to reproduce a comprehensive review, but to organize the main theoretical lessons around a few representative examples. We then present a new high-energy probe of CLFV using cosmic-ray observations at neutrino telescopes~\cite{Maitra:2026qhi}.

\begin{table*}[h!]
\centering
\caption{Current upper limits on the BRs of selected CLFV processes (updated from Ref.~\cite{Ardu:2022sbt}) and future sensitivities. See Ref.~\cite{ParticleDataGroup:2026mpi} for details.}
\label{tab:clfv_limits_updated}
\vspace{0.2cm}
\tiny
\setlength{\tabcolsep}{5pt}
\begin{tabular}{lllll}
\hline
Process & Current limit & Experiment & C.L. & Future sensitivity (experiment)\\
\hline
$\mu^+\to e^+\gamma$ & $1.5\times10^{-13}$ & MEG II & 90\% & $\sim6\times10^{-14}$ (MEG II design goal)\\
$\mu^+\to e^+e^-e^+$ & $1.0\times10^{-12}$ & SINDRUM & 90\% & $\sim10^{-15}$--$10^{-16}$ (Mu3e)\\
$\mu^-{\rm Ti}\to e^-{\rm Ti}$ & $4.3\times10^{-12}$ & SINDRUM II & 90\% & --\\
$\mu^-{\rm Au}\to e^-{\rm Au}$ & $7.0\times10^{-13}$ & SINDRUM II & 90\% & --\\
$\mu^-{\rm Al}\to e^-{\rm Al}$ & -- & -- & -- & $\sim10^{-16}$--$10^{-17}$ (Mu2e/COMET)\\
$\mu^-N\to e^+N'$ & $5.7\times10^{-13}$ & SINDRUM II & 90\% & --\\
\hline
$\tau^\pm\to e^\pm\gamma$ & $3.3\times10^{-8}$ & BaBar & 90\% & $\mathcal{O}(10^{-9})$ (Belle II)\\
$\tau^\pm\to\mu^\pm\gamma$ & $4.2\times10^{-8}$ & Belle/BaBar & 90\% & $\mathcal{O}(10^{-9})$ (Belle II)\\
$\tau\to eee$ & $2.7\times10^{-8}$ & Belle & 90\% & $\mathcal{O}(10^{-10})$ (Belle II)\\
$\tau\to\mu\mu\mu$ & $2.1\times10^{-8}$ & Belle & 90\% & $\mathcal{O}(10^{-10})$ (Belle II)\\
$\tau\to\mu ee$ & $1.8\times10^{-8}$ & Belle & 90\% & $\mathcal{O}(10^{-10})$ (Belle II)\\
$\tau\to e\mu\mu$ & $2.7\times10^{-8}$ & Belle & 90\% & $\mathcal{O}(10^{-10})$ (Belle II)\\
$\tau\to\pi^0e$ & $8.0\times10^{-8}$ & Belle & 90\% & $\mathcal{O}(10^{-10})$ (Belle II)\\
$\tau\to\pi^0\mu$ & $1.1\times10^{-7}$ & BaBar & 90\% & $\mathcal{O}(10^{-10})$ (Belle II)\\
$\tau\to\eta e$ & $9.2\times10^{-8}$ & Belle & 90\% & $\mathcal{O}(10^{-10})$ (Belle II)\\
$\tau\to\eta\mu$ & $6.5\times10^{-8}$ & Belle & 90\% & $\mathcal{O}(10^{-10})$ (Belle II)\\
$\tau\to\rho^0e$ & $1.8\times10^{-8}$ & Belle & 90\% & $\mathcal{O}(10^{-10})$ (Belle II)\\
$\tau\to\rho^0\mu$ & $1.2\times10^{-8}$ & Belle & 90\% & $\mathcal{O}(10^{-10})$ (Belle II)\\
\hline
$\pi^0\to\mu^\pm e^\mp$ & $3.8\times 10^{-10}$ & KTeV & 90\% & --\\
$K_L^0\to\pi^0\mu^\pm e^\mp$ & $7.6\times10^{-11}$ & KTeV & 90\% & --\\
$K_L^0\to\mu^\pm e^\mp$ & $4.7\times10^{-12}$ & BNL E871 & 90\% & ${\cal O}(10^{-13})$ (HIKE)\\
$K^+\to\pi^+\mu^+e^-$ & $1.3\times10^{-11}$ & BNL E865 & 90\% & --\\
\hline
$J/\psi\to\mu^\pm e^\mp$ & $1.5\times10^{-7}$ & BESIII & 90\% & ${\cal O}(10^{-11})$ (STCF) \\
$J/\psi\to\tau^\pm e^\mp$ & $7.5\times10^{-8}$ & BESIII & 90\% & ${\cal O}(10^{-9})$ (STCF)\\
$J/\psi\to\tau^\pm\mu^\mp$ & $2.0\times10^{-6}$ & BESIII & 90\% & ${\cal O}(10^{-9})$ (STCF) \\
$B^0\to\mu^\pm e^\mp$ & $2.8\times10^{-9}$ & LHCb & 90\% & ${\cal O}(10^{-10})$ (LHCb upgrade/Belle II)\\
$B^0\to\tau^\pm e^\mp$ & $2.8\times10^{-5}$ & BaBar & 90\% & ${\cal O}(10^{-8})$ (Belle II)\\
$B^0\to\tau^\pm\mu^\mp$ & $1.4\times10^{-5}$ & LHCb & 90\% & ${\cal O}(10^{-8})$ (Belle II) \\
$B\to K\mu^\pm e^\mp$ & $7.0\times10^{-9}$ & LHCb & 90\% & ${\cal O}(10^{-10})$ (LHCb upgrade)\\
$B\to K^*\mu^\pm e^\mp$ & $1.8\times10^{-8}$ & LHCb & 90\% & ${\cal O}(10^{-10})$ (LHCb upgrade)\\
$B^+\to K^+\tau^\pm e^\mp$ & $1.5\times10^{-5}$ & LHCb & 90\% & ${\cal O}(10^{-7})$ (Belle II)\\
$B^+\to K^+\tau^\pm\mu^\mp$ & $3.9\times10^{-5}$ & LHCb & 90\% & ${\cal O}(10^{-7})$ (Belle II)\\
$B_s^0\to\mu^\pm e^\mp$ & $1.1\times10^{-8}$ & LHCb & 90\% & ${\cal O}(10^{-9})$ (LHCb upgrade)\\
$B_s^0\to\tau^\pm\mu^\mp$ & $3.4\times10^{-5}$ & LHCb & 90\% & ${\cal O}(10^{-6})$ (LHCb upgrade)\\
\hline
$Z\to\mu^\pm e^\mp$ & $7.5\times10^{-7}$ & ATLAS & 95\% & ${\cal O}(10^{-10})$ (FCC-ee)\\
$Z\to\tau^\pm e^\mp$ & $5.0\times10^{-6}$ & ATLAS & 95\% & ${\cal O}(10^{-9})$ (FCC-ee)\\
$Z\to\tau^\pm\mu^\mp$ & $6.5\times10^{-7}$ & ATLAS & 95\% & ${\cal O}(10^{-9})$ (FCC-ee)\\
$h\to\mu^\pm e^\mp$ & $3.0\times10^{-5}$ & CMS & 95\% & ${\cal O}(10^{-7})$ (FCC-ee)\\
$h\to\tau^\pm e^\mp$ & $1.6\times10^{-3}$ & ATLAS & 95\% & ${\cal O}(10^{-5})$ (FCC-ee)\\
$h\to\tau^\pm\mu^\mp$ & $1.5\times10^{-3}$ & CMS & 95\% & ${\cal O}(10^{-5})$ (FCC-ee)\\
\hline
\end{tabular}
\end{table*}

\section{Experimental Progress}

The experimental history of CLFV searches illustrates the remarkable progress made over the past decades~\cite{kaon19}, with limits improving steadily from the early cosmic ray muon experiments through pion and muon beams, while also showing the growing complementary role of LFV decay channels involving heavier mesons, $\tau$, $Z$, and Higgs bosons at high-energy colliders. Current limits and future sensitivities for selected CLFV processes are summarized in Table~\ref{tab:clfv_limits_updated}.

\section{From a CLFV signal to the underlying theory}
A central difficulty in CLFV phenomenology is that low-energy experiments usually do not produce the new particle directly. Instead, they probe virtual effects. Different microscopic theories can therefore lead to similar low-energy rates. A single measurement can rarely determine whether the underlying model is, for example, a seesaw, a supersymmetric theory, a leptoquark model, an extra scalar, or a new vector boson.

The correct strategy is to combine observables with different operator dependence. Correlations among $\mu\to e\gamma$, $\mu\to3e$, $\mu N\to eN$, $\tau$ decays, and LFV Higgs or $Z$ decays can reveal the flavor structure and chirality of the effective interaction. The situation becomes even richer once leptonic CP phases are allowed: cancellations or enhancements can weaken otherwise simple correlations. In such cases complementary collider and low-energy measurements are essential.

This lesson can be formalized in effective field theory (EFT). Below the mass scale of new particles, the Lagrangian can be organized as
\begin{equation}
\mathcal L_{\rm SMEFT}=\mathcal L_{\rm SM}
+\frac{1}{\Lambda}\sum_i C_i^{(5)}Q_i^{(5)}
+\frac{1}{\Lambda^2}\sum_i C_i^{(6)}Q_i^{(6)}+\cdots .
\end{equation}
The unique dimension-five Weinberg operator does not produce observable CLFV at tree level since it is suppressed by neutrino mass. The leading general CLFV effects therefore arise at dimension six.

The relevant operator classes include dipoles,
\begin{equation}
Q_{eB}=(\bar L\sigma^{\mu\nu}e_R)HB_{\mu\nu},\qquad
Q_{eW}=(\bar L\sigma^{\mu\nu}\tau^I e_R)HW^I_{\mu\nu},
\end{equation}
four-lepton operators such as
\begin{equation}
Q_{\ell\ell}=(\bar L\gamma_\mu L)(\bar L\gamma^\mu L),\qquad
Q_{ee}=(\bar e_R\gamma_\mu e_R)(\bar e_R\gamma^\mu e_R),
\end{equation}
and semileptonic operators,
\begin{equation}
Q_{\ell q}^{(1,3)}=(\bar L\gamma_\mu\tau^{I}L)(\bar Q\gamma^\mu\tau^{I}Q),
\end{equation}
with analogous operators involving $u_R$ and $d_R$~\cite{Crivellin:2013hpa}. Higgs-current operators modify LFV $Z$ couplings and can connect low-energy observables to collider searches~\cite{Calibbi:2021pyh}.

The complete dimension-six basis relevant for lepton-flavor observables was analyzed in Ref.~\cite{Crivellin:2013hpa}. The essential point is that a single experimental observable constrains only a direction in Wilson-coefficient space. Combining radiative decay, three-body decay and conversion measurements rotates the allowed directions and can dramatically improve model discrimination~\cite{Calibbi:2017uvl}.

The $\mu$--$e$ sector is presently the most sensitive. The radiative decay $\mu\to e\gamma$ probes the electromagnetic dipole operator and is especially clean experimentally. The decay $\mu\to3e$ adds sensitivity to contact interactions and off-shell photon penguins. Coherent conversion, $\mu N\to eN$, is sensitive to both dipole and semileptonic interactions and benefits from coherent enhancement over the nucleons in a nucleus. The combination of these channels is therefore much more informative than any single limit.

\section{Representative ultraviolet realizations}
\subsection{Heavy neutral leptons}
Heavy neutral leptons (HNLs) provide perhaps the simplest illustration of how a new particle can remove the extreme SM suppression. In a generic type-I seesaw-like setup,
\begin{equation}
\mathcal L\supset -Y_\nu\,\bar L\widetilde H \nu_R-
\frac12\overline{\nu_R^c}M_R\nu_R+\mathrm{h.c.},
\end{equation}
and after electroweak symmetry breaking the light-neutrino mass matrix is
\begin{equation}
m_\nu\simeq-\frac{v^2}{2}Y_\nu M_R^{-1}Y_\nu^T.
\end{equation}
The charged-current mixing matrix is no longer exactly unitary once heavy states are retained. This changes the cancellation structure of the loop amplitude. For $\mu\to e\gamma$, for example, we get~\cite{Alonso:2012ji}
\begin{equation}
{\rm BR}(\mu\to e\gamma)
=\frac{3\alpha}{32\pi}
\frac{\left|\sum_k U_{\mu k}^*U_{e k}F(x_k)\right|^2}
{(UU^\dagger)_{\mu\mu}(UU^\dagger)_{ee}},
\qquad x_k=\frac{m_k^2}{m_W^2},
\label{eq:seesaw}
\end{equation}
where the loop function $F(x_k)$ approaches a constant plus corrections at large $x_k$. For the three light neutrinos, unitarity enforces the GIM cancellation and the remaining amplitude is proportional to neutrino mass splittings. For heavy states the cancellation is incomplete because the enlarged mixing matrix contains nonunitarity effects.

The price is the tension with neutrino-mass generation. Large active-sterile mixing and relatively light HNLs are required for better CLFV sensitivity, while the observed small neutrino masses typically prefer smaller mixing unless additional structure or approximate symmetries are present. This is a recurring theme in CLFV: observable rates often require a technically motivated mechanism, such as the inverse seesaw, that separates the origin of small neutrino masses from the size of active-sterile mixing and, consequently, from the CLFV rate.

\subsection{Type-II seesaw}
The type-II seesaw introduces a scalar $SU(2)_L$ triplet $\Delta$ carrying hypercharge and coupled to lepton doublets as
\begin{equation}
\mathcal L\supset -\frac12 Y_{\Delta}^{ij}
L_i^T C i\tau_2\Delta L_j+\mathrm{h.c.}
\end{equation}
After the triplet acquires a small vacuum expectation value, $v_\Delta$, the neutrino mass matrix is directly related to the Yukawa matrix,
\begin{equation}
m_\nu=\sqrt{2}\,Y_\Delta v_\Delta.
\end{equation}
This direct relation makes the model especially predictive for flavor observables.

A characteristic feature is the presence of tree-level CLFV mediated by the neutral triplet component. In particular, $\mu\to3e$ can be generated at tree level, while $\mu\to e\gamma$ first appears at one loop. Consequently, $\mu\to3e$ often dominates in regions of parameter space where the triplet couplings are responsible for neutrino masses. The same couplings govern decays of the doubly charged scalar, creating a striking connection between low-energy flavor violation and direct searches for $\Delta^{\pm\pm}$ at colliders. The correlations are sensitive to the neutrino mass ordering and to the absolute neutrino mass scale~\cite{Dev:2018tox}.

The type-II example demonstrates why several CLFV channels should be pursued simultaneously. A future observation of $\mu\to3e$ without a corresponding signal in $\mu\to e\gamma$ would point toward non-dipole interactions and could favor models of this type over a purely dipole-dominated scenario.

\subsection{Radiative neutrino mass models}
Radiative neutrino-mass models provide another instructive class. In the Zee model, an additional charged scalar and a second Higgs doublet generate neutrino masses at one loop. The same flavor couplings induce CLFV. An important feature is that neutral Higgs exchange can induce tree-level CLFV, while charged-Higgs contributions arise at loop level. The mass spectrum can consequently allow the charged scalar to be much lighter than the neutral scalar, while yielding sizable CLFV rates.

These models also connect charged-lepton flavor to neutrino non-standard interactions (NSI)~\cite{Babu:2019mfe}. Thus, a viable radiative neutrino-mass model can simultaneously leave signatures in CLFV and in neutrino scattering, providing a useful example of the broader complementarity between apparently different probes.

The scotogenic model offers another particularly clean realization in which neutrino masses arise radiatively and a discrete symmetry stabilizes a dark-matter candidate. CLFV proceeds through loops containing the inert scalar doublet and singlet fermions. Although $\ell_i\to\ell_j\gamma$ is often used as the canonical constraint, complete calculations show that $\ell_i\to3\ell_j$ and $\mu$--$e$ conversion can be competitive or stronger in relevant regions of parameter space~\cite{Toma:2013zsa}. Thus even in a loop-induced model, the operator pattern need not be purely dipole-like.

\section{Correlations, CP phases, and collider complementarity}
Even after specifying a class of ultraviolet models, flavor observables can exhibit substantial correlations and cancellations. For example, in seesaw scenarios the rates of different CLFV channels depend on products of mixing matrices [cf.~Eq.~(\ref{eq:seesaw})]. In simple limits these products can lead to narrow correlations in the space of branching ratios. Once leptonic CP phases are varied, however, interference among contributions can destroy a naive one-to-one mapping.

This issue is particularly important for LFV Higgs and $Z$ decays. A future collider may probe $h\to\mu\tau$, $h\to e\tau$, or $Z\to\ell_i\ell_j$ at levels inaccessible to low-energy experiments, while low-energy measurements can constrain the same EFT coefficients indirectly. The analysis of Ref.~\cite{Abada:2022asx} emphasizes that leptonic CP phases can lead to a loss of simple correlations between LFV Higgs and $Z$ observables. The practical consequence is that complementary probes are not optional: they are necessary to reconstruct the flavor structure.

The EFT viewpoint makes the complementarity transparent. A four-lepton operator contributes strongly to $\mu\to3e$ but may have only an indirect effect on a collider observable. A semileptonic operator can be much more efficiently constrained by coherent conversion or high-energy scattering. Dipole operators dominate radiative decays. LFV Higgs and $Z$ decays probe operators involving the Higgs current or electroweak gauge bosons. The pattern of measurements therefore provides a tomography of the effective interaction.

Future lepton colliders are particularly interesting because of their enormous samples of electroweak bosons. A Tera-$Z$ facility could collect approximately $10^{12}$ $Z$ decays, opening sensitivity to LFV $Z$ decays at levels far beyond existing experiments. Similarly, a future Higgs factory can probe the Higgs LFV decays to an unprecedented level.  At the same time, high-energy colliders directly test the mediators that generate the low-energy operators. For instance, the purported CLFV signal observed by  CMS~\cite{CMS:2023pte}, if confirmed, may point to a leptophilic two-Higgs-doublet model~\cite{Afik:2023vyl}.  This is why the most robust CLFV program must combine  intensity-frontier experiments, flavor factories, and high-energy colliders rather than choosing one in isolation.

\section{A new idea: CLFV at neutrino telescopes}
The discussion so far concerns conventional CLFV observables. A qualitatively different opportunity arises in neutrino telescopes. Large Cherenkov detectors are normally viewed as neutrino detectors, but they also function as enormous charged-lepton detectors. Cosmic rays interacting in the atmosphere generate energetic muons, and a substantial fraction of these muons penetrate deep into ice or water. Their abundance turns what is normally a background into a potential source of new physics.
\subsection{Why is $\mu\to \tau$ conversion special?}

\begin{wrapfigure}{r}{0.3\textwidth}
    \centering
    \includegraphics[width=\linewidth]{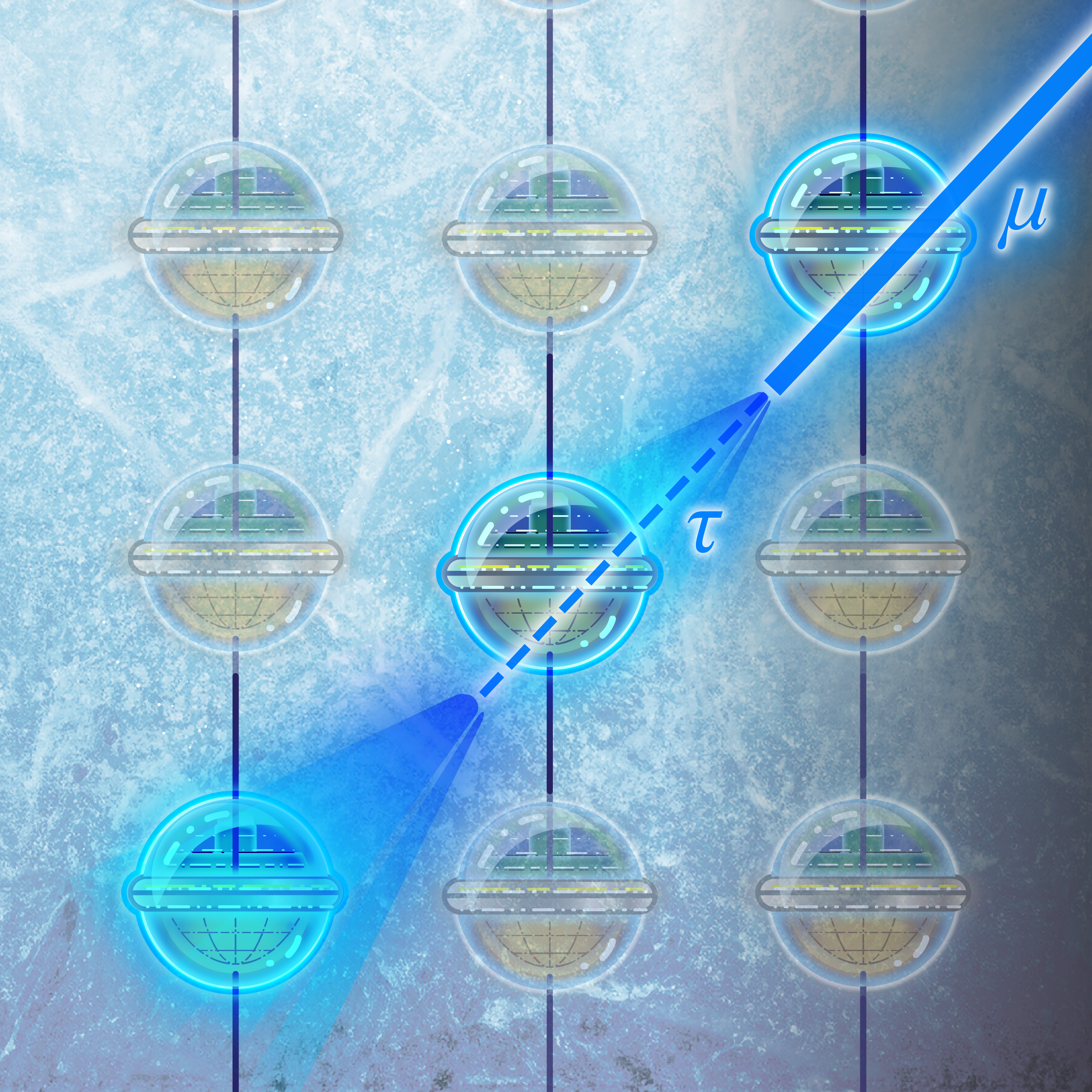}
    \caption{Schematic illustration of the proposed CLFV signature at a neutrino telescope: a muon track transitioning into a tau track, which subsequently initiates a hadronic shower within the detector~\cite{Maitra:2026qhi}.}
    \label{fig:Signal}
\end{wrapfigure}

The central proposal of Ref.~\cite{Maitra:2026qhi} is to search for 
$\mu^{\pm}N\rightarrow \tau^{\pm}N$ 
inside a neutrino telescope. The incoming muon leaves a track; at the CLFV interaction point it converts into a tau, which subsequently propagates a finite distance and decays, often producing a hadronic cascade. The topology is therefore a track-to-track-plus-cascade transition (see Fig.~\ref{fig:Signal}),  rather than the prompt electromagnetic activity expected for a $\mu\to e$ conversion. This makes the $\mu\to\tau$ channel comparatively distinctive in a Cherenkov detector.

The idea has an important conceptual advantage. Low-energy experiments provide extremely strong constraints on $\mu\to e$ flavor violation (cf.~Table~\ref{tab:clfv_limits_updated}), while $\mu\to\tau$ conversion on nuclei is much less constrained because it is difficult to realize with sufficient sensitivity in conventional fixed-target setups. The reason is that muons can be stopped and stored in enormous numbers, whereas a low-energy experiment cannot easily create an intense, controlled beam of muons energetic enough to produce a tau through flavor conversion. The neutrino-telescope environment naturally supplies both the incoming high-energy muon beam and a large detector volume. The resulting probe therefore targets a comparatively unexplored direction in flavor space.

The tau itself is also an asset. At TeV energies its boosted decay length can be substantial on detector scales. Hadronic tau decays produce compact cascades spatially separated from the incoming track. This topology is qualitatively different from ordinary continuous ionization of a muon and from a purely electromagnetic shower. Thus the same tau instability that makes the channel difficult in a conventional low-energy experiment creates a useful signature in a large-volume Cherenkov detector.

The relevant input is the measured distribution of cosmic-ray-induced muons. IceCube's 12-year published data contains 792 billion  downgoing cosmic-ray muon events~\cite{IceCube:2024pnx}, an exceptionally large sample for a BSM search. 
This measured flux is the starting point for our CLFV event-rate calculation. A $\mu-\tau$ CLFV interaction changes the flavor of the propagating muon. The expected number of tau events can be written schematically as a convolution of three ingredients: the incident muon flux, propagation and energy loss of the muon before the interaction, and the probability that the produced tau is reconstructed in the detector.

The dominant irreducible background in the conceptual analysis is a through-going muon whose stochastic energy loss produces a topology resembling the tau cascade. The proposed signal differs because the charged track changes identity at an interaction point and the resulting tau has a finite propagation distance before decay. A realistic experimental analysis would require detector-level event reconstruction, muon stochastic energy-loss modeling, tau-tagging efficiencies, and a full treatment of atmospheric backgrounds and detector systematics.

\subsection{Model-independent EFT interpretation}
The cleanest way to quantify the new probe is to begin with semileptonic operators that mediate $\mu q\to\tau q$. At the scale relevant to the detector, one may write
\begin{equation}
\mathcal L_{\rm EFT}\supset
\frac{1}{\Lambda^2}\sum_{\Gamma}
C_6^\Gamma(\bar\mu\Gamma\tau)(\bar q\Gamma q),
\end{equation}
where $\Gamma$ labels scalar, pseudoscalar, vector, and axial-vector Lorentz structures. We set the Wilson coefficients $C_6$ to order unity when quoting the corresponding constraints on the new-physics scale $\Lambda$.

The conversion probability is proportional to the partonic or nucleon-level cross section and hence approximately scales as $\Lambda^{-4}$. Because the incoming flux is enormous, even a small conversion probability can produce an appreciable number of events. The resulting sensitivity is therefore complementary to conventional decay searches, whose rates depend on different combinations of couplings and mediator masses.

\begin{figure*}[t!]
\centering
    \includegraphics[width=\linewidth]{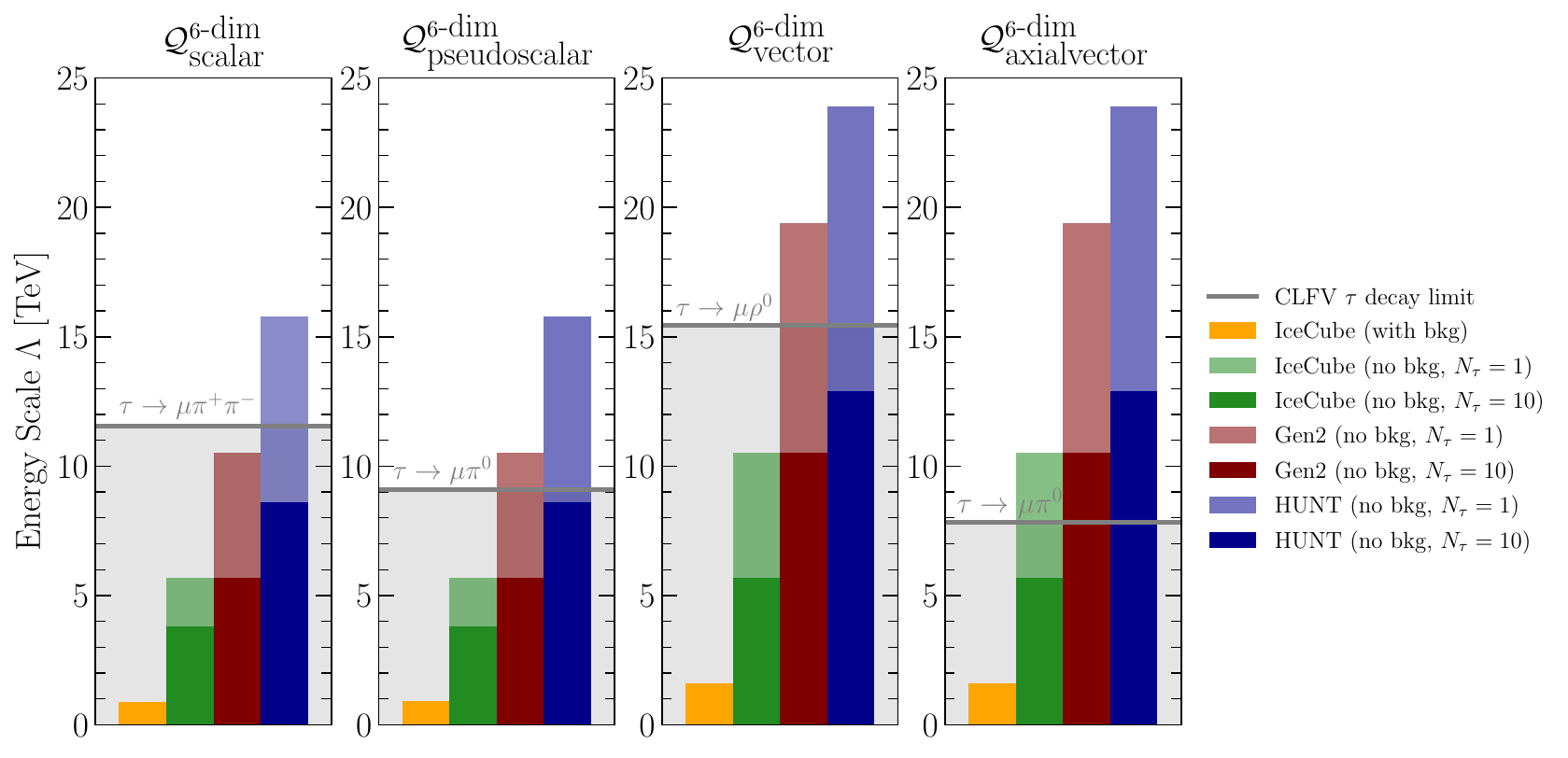}
    \caption{The 95\% C.L. sensitivity to the new-physics scale \(\Lambda\) for various EFT operators~\cite{Maitra:2026qhi}. The gray horizontal lines indicate the existing constraints from low-energy \(\tau\)-LFV observables.}
    \label{fig:6dim}
\end{figure*}
Figure~\ref{fig:6dim} summarizes our results. Existing CLFV decay limits provide a baseline, while the IceCube estimate with backgrounds and idealized background-free projections show the complementarity. The next-generation detectors, such as IceCube-Gen2 and HUNT, can push the scale substantially higher, particularly for vector and axial-vector interactions, thereby probing new parameter space. The exact numerical reach depends on the event selection and the assumed number of tolerated background events. The sensitivity curves shown here include two conceptually different assumptions: an estimate with an irreducible background and idealized projections in which one or ten signal events are required in the absence of background. The latter should be interpreted as a measure of ultimate statistical reach, not as a realistic forecast. A future experimental paper would need to establish a background model and a likelihood using control regions. Nevertheless, these projections are valuable because they show that the physics opportunity is not limited by the total number of muons.

\subsection{Connection to $Z$ and Higgs LFV}
The same flavor structure can also induce LFV decays of the $Z$ and Higgs bosons. We find that in the $Z$ case, the neutrino telescope sensitivities can be competitive with the existing collider limits on BR($Z\to\mu\tau$), while in the Higgs case, the IceCube sensitivity is not as competitive primarily due to the suppressed Higgs Yukawa coupling to first-generation quarks.

However, the neutrino-telescope sensitivity should not be interpreted as a direct BR measurement in a model-independent way: it constrains the CLFV scattering interaction $\mu N\to \tau N$, which can be related to $Z$ or Higgs observables only within a specified ultraviolet model. This distinction is important when combining constraints.

\section{Experimental outlook}
The future CLFV program is unusually broad (cf.~Table~\ref{tab:clfv_limits_updated}). In the $\mu$--$e$ sector, MEG-II, Mu3e, Mu2e and COMET target radiative decay, three-body decay and coherent conversion with improvements of several orders of magnitude. In the $\tau$ sector, Belle II and future flavor factories probe a large collection of decay modes. At colliders, LFV Higgs and $Z$ searches extend sensitivity to higher scales and to operator structures not easily isolated at low energy.

Neutrino telescopes add a qualitatively new ingredient: a huge sample of high-energy charged leptons traversing a macroscopic target. The first implementation uses the existing IceCube cosmic-ray muon data, but the concept is not limited to IceCube. IceCube-Gen2 will offer larger instrumented volumes, while other large water-Cherenkov detectors such as KM3NeT, Baikal-GVD, P-ONE, TRIDENT and HUNT, could in principle implement analogous searches. The method also suggests a broader philosophy: backgrounds in one analysis can become signal samples for another if their kinematics and topology are sufficiently well understood.

The key experimental developments needed to turn the proposal into a dedicated search are clear. First, a detector-level simulation should be used to model the track-to-tau topology, including stochastic muon losses and tau decay. Second, the dominant backgrounds should be estimated from data-driven control samples. Third, a multivariate or topology-based tau classifier should be developed. Finally, systematic uncertainties in the atmospheric muon flux and detector response should be propagated into the CLFV likelihood.

A particularly attractive feature is the possibility of a background-controlled analysis. If the tau conversion point and subsequent cascade can be reconstructed with sufficient fidelity, one can exploit the characteristic spatial separation between the incoming track and the tau decay. This may suppress conventional stochastic-loss backgrounds substantially. The ultimate sensitivity will therefore be determined not only by exposure but also by the quality of tau reconstruction.

A useful long-term goal would be a combined analysis across multiple telescopes. Independent detectors have different geometries, optical properties, overburdens and atmospheric muon systematics. A common CLFV signal should scale predictably with exposure and target material. A combined observation would thus provide an important cross-check and could help distinguish new physics from an unmodeled detector effect.

\section{Broader theoretical implications}
CLFV should not be viewed as an isolated flavor observable. It sits at the intersection of several major questions in particle physics. In seesaw models it is connected to neutrino mass and mixing. In radiative models it can correlate with dark matter and neutrino non-standard interactions. In Higgs-sector models it probes the structure of electroweak symmetry breaking. In models with new gauge bosons it can be related to anomaly cancellation and flavor symmetries.

This broader perspective also explains why a null result still remains informative. If the CLFV scale is pushed upward, simple weak-scale models with anarchic flavor couplings become increasingly constrained. Viable theories then require structure: approximate symmetries, alignment, decoupling, radiative suppression, or special flavor textures. Conversely, an observed pattern across multiple channels could reveal that structure. For example, a dipole-dominated signal would suggest a very different microscopic origin from a simultaneous enhancement of $\mu\to3e$ and $\mu N\to eN$ with a suppressed $\mu\to e\gamma$ rate.

The same logic applies to the proposed $\mu\to\tau$ conversion search. An observation would immediately imply BSM physics, but interpreting it would require combining the neutrino-telescope result with $\tau$ decays, LFV $Z$ and Higgs searches, electroweak precision measurements, and direct mediator searches. The scientific payoff is that one would gain access to a flavor transition that is currently much less constrained than the $\mu$--$e$ sector.

\section{Conclusions}
CLFV is among the cleanest probes of BSM physics. The accidental flavor symmetry of the charged-lepton sector is broken by neutrino masses, but the minimal neutrino-induced CLFV rates are so tiny that any observable signal would reveal new physics. This makes CLFV searches  uniquely powerful. 

The broader lesson is that no single experiment is likely to reveal the full story, but the combination of low-energy precision, collider production, neutrino scattering, and cosmic-ray data can provide a remarkably complete map of new flavor dynamics.

\section*{Acknowledgments}
This work was
partly supported by the U.S. Department of Energy under
grant No. DE-SC0017987 and by a Humboldt Fellowship from the Alexander von Humboldt Foundation.

\bibliographystyle{ieeetr}
\bibliography{refs}
\end{document}